%% file: GEOpolarimetry0.tex
\documentclass[journal,12pt,draftclsnofoot,onecolumn]{IEEEtran}

\usepackage{cite}
\usepackage[dvips]{graphicx}
\usepackage{multicol}

\usepackage{amsfonts}
\usepackage{amssymb}
\usepackage{amsmath}
\usepackage{floatflt}

\begin{document}
\title{On mathematical and physical principles of transformations of the coherent radar backscatter matrix}
\author{David~Bebbington~\IEEEmembership{} and~Laura~Carrea~\IEEEmembership{}
\thanks{This work was supported by the Marie Curie Research Training
Network AMPER (Contract number HPRN-CT-2002-00205).}
\thanks{D. Bebbington and L. Carrea are with the Centre for Remote Sensing \& Environmetrics, School of Computer Science and Electronic Engineering, University of Essex, U.K. (email: david@essex.ac.uk;
lcarrea@essex.ac.uk).}}

\maketitle

\begin{abstract}
The congruential rule advanced by Graves for polarization basis transformation of the radar backscatter matrix is now often misinterpreted as an example of consimilarity transformation. However, consimilarity transformations imply a physically unrealistic antilinear time-reversal operation. This is just one of the approaches found in literature to the description of transformations where the role of conjugation has been misunderstood. In this paper, the different approaches are examined in particular in respect to the role of conjugation. In order to justify and correctly derive the congruential rule for polarization basis transformation and properly place the role of conjugation, the origin of the problem is traced back to the derivation of the antenna hight from the transmitted field. In fact, careful consideration of the role played by the Green's dyadic operator relating the antenna height to the transmitted field shows that, under general unitary basis transformation, it is not justified to assume a scalar relationship between them.  Invariance of the voltage equation shows that antenna states and wave states must in fact lie in dual spaces, a distinction not captured in conventional Jones vector formalism. Introducing spinor formalism, and with the use of an alternate spin frame for the transmitted field a mathematically consistent implementation of the directional wave formalism is obtained.  Examples are given comparing the wider generality of the congruential rule in both active and passive transformations with the consimilarity rule.
\end{abstract}


\IEEEpeerreviewmaketitle

\section{Introduction}
\input{GP0Introduction}

\section{Defining relations in radar polarimetry} \label{relations}
\input{GP0Relations}

\section{A critique of former approaches}\label{critique}
\input{GP0Critique}

\section{A linear theory of Graves' rule} \label{Graves}
\input{GP0Graves}

\section{Examples and results} \label{Examples}
\input{GP0Examples}

\section{Conclusions}
\input{GP0Conclusion}

\appendices
\section{The priming operation and linear transformations}
\input{GP0Appendix}

\section*{Acknowledgment}
The work on Geometric Polarization was supported by the U.S.
Office of Naval Research and the European Union (Marie Curie
Research Training Network RTN AMPER, Contract number
HPRN-CT-2002-00205). The authors also owe a debt of gratitude in particular
to Prof. Wolfgang Boerner and the late Dr. Ernst L\"{u}neburg, who
were prime movers in stimulating the debate on the foundations of
polarimetry, regardless of where that might lead.
Finally, the authors would like to thank the anonymous reviewers
for their valuable comments and suggestions to improve the
quality of the paper.

\bibliographystyle{IEEEtran}
\bibliography{IEEEabrv,bibliografia}

\vspace{-1cm}
\begin{biography}[{\includegraphics[width=1in,height=1.25in,clip,keepaspectratio]{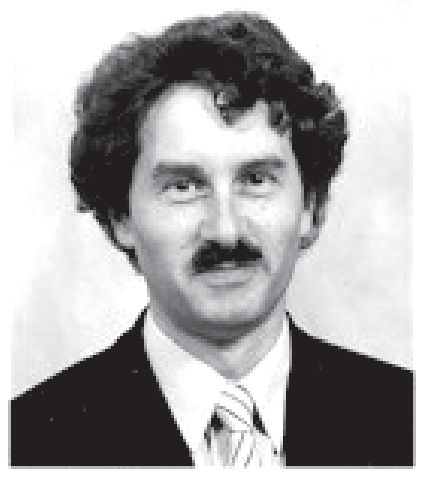}}]{David Bebbington}
Received the B.A. degree in experimental and theoretical physics
and the Ph.D. degree in radio astronomy from Cambridge University,
Cambridge, U.K., in 1977 and 1986, respectively. From 1981 to
1984, he worked on millimeter wave propagation research at the
Rutherford Appleton Laboratory. Since 1984, he has been at the
University of Essex, Essex, U.K., and currently holds the post of
Senior Lecturer in the Department of Computing and Electronic
System. His interests are polarimetry, applications of wave
propagation in remote sensing, and weather radars.
\end{biography}

\vspace{-1cm}

\begin{biography}[{\includegraphics[width=1in,height=1.25in,clip,keepaspectratio]{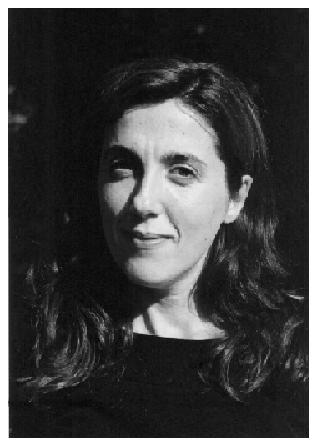}}]{Laura Carrea}
received the Laurea degree in physics from the University of
Turin, Italy in 1997. From 1999 to 2004 she worked at the Faculty
of Electrical Engineering and Information Technology at the
Chemnitz University of Technology in Germany partly with a
Fellowship in the frame of the Marie Curie TMR Network "Radar
Polarimetry: Theory and Application". Since 2004 she is with
Centre for Remote Sensing \& Environmetrics, Department of
Computing and Electronic System, University of Essex, U.K. Her
main interests center on radar polarimetry.
\end{biography}

\vspace{-0.8cm}

\end{document}

%% file: GP0Introduction.tex

Apart from the fact that it is an important principle of mathematical physics that formulae involving vectors should be expressible in a way that is independent of the basis, there are often practical reasons to transform bases. Sometimes simplicity is achieved in a particular basis, especially if the symmetry of a problem favours a particular representation, or that in some representation a convenient parametrization arises that best captures the phenomena of interest. Recent examples in polarimetry include the following \cite{galletti2008}, \cite{galletti2011}, \cite{paladini}.

This paper should be regarded as the precursor to a
planned series on the foundations of polarimetry. Our initial plan is to clarify some long standing ambiguities and conflicting viewpoints expressed in the previous literature and to motivate a new approach. Such an approach will redevelop the foundation of polarimetry that is consistent with physical principles and exhibits a coherent mathematical framework.
In particular, our purpose here is not a general review, but to focus instead on a
small number of key contributions to the literature where we can
most clearly demonstrate that ambiguities have been introduced in the process of justifying the current formulation of polarimetry. More than any other concept in
polarimetry, it seems that the equivalence relation for
polarization states upon wave reversal remains problematic (e.g. \cite{luneburgreply}, \cite{hubbertreply}).
Previous papers addressing the foundations of polarimetry in the
context of remote sensing radar have had to run the gauntlet over
the status of the so-called backscatter alignment convention (BSA). This convention
is almost universally adopted in monostatic radar, and is
associated with this issue of wave reversal that has proved to be
so contentious historically \cite{kostinski}, \cite{mieras},
\cite{kostinskiR}, \cite{hubbertbringi}, \cite{luneburgreply},
\cite{hubbertreply}. So, although this was not our intended
starting point in reforming the foundations of polarimetry, it
seems that we cannot set out our stall until certain fundamentals
regarding this question are dealt with. It is beyond the scope of this paper to comprehensively introduce our formal methods. However, our intention is to show how the problem can be addressed, employing some of the ideas that flow from a new perspective.

The solution to the problem of how correctly to account for Graves' congruential rule \cite{graves}, as it applies to the transmitted and received polarization states of the electric fields is developed in this paper in terms of spinors. References to spinors as descriptors of coherent polarization states have appeared previously in the literature \cite{cloude:liegroups} \cite{cloude:vector} \cite{torres}, and more recently, applications of quaternion algebra to scattering matrices have been presented \cite{carrea:igarss}, \cite{souyris}. To date, however, the fundamental significance of spinors in polarization representation has not been discussed in any depth.
Given that within the polarimetric community there is generally much less familiarity with spinor algebra than with conventional vector algebra, we have to consider ourselves under some duty to make a case for it. All other things being equal, one could say rather, why leave spinors out of the picture?  Spinors are the carriers for the unitary group SU(2), which is what is used to make basis changes. All conventional polarimetry theory refers at least implicitly to SU(2), and spinors are what that group operates on. In a strict sense, if one consider Jones vectors as vectors rather than spinors, problems are already lining up: do we consider Euclidean or Hermitian products as fundamental?  For orthogonality tests, and gauging of intensity, one traditionally uses the Hermitian inner product; in the voltage equation where the response of an antenna to a received field is evaluated, however, it appears that a conventional Euclidean inner product is taken.  The essence of inner products is that they should be invariant under basis change - but in terms of conventional forms of vector algebra one needs to ask how both could be true if each is only invariant under distinct and incompatible group actions.  The compelling answer to these and other curiosities in polarimetry that have puzzled several generations of polarimetrists is that the conventional vector algebra does not have the capability to express the different complexions of \emph{vectorial} objects that appear in polarimetry.  Spinor algebra (with indices), on the other hand, expresses succinctly exactly the four different forms of 2-vector that are required.  Firstly, as this paper demonstrates, in addition to gradient-like objects (the fields) belonging to the space of covariant fields, we need to have coordinate-like objects such as complex antenna height.  For each of these we need a related, conjugated space.  Conjugate spaces have antilinear transformational properties, and therefore require a separate spaces of representation from \emph{normal} linear space.  As this paper will show, it has been a long standing and unfortunate  misconception in polarimetry that the backscatter alignment convention (BSA) implies that backscattered waves are expressed in a conjugate representation. This would imply that backscatter is antilinear in character.  Such an interpretation leads to a number of absurd conclusions involving \emph{time reversal}, which are not warranted by the physics.  To be clear, conjugate representations are required in operations relating to correlation.  Here the time reversal implied is conceptually valid, because it is accounted for in the autocorrelation kernel of which the complex correlation is the Fourier transform.
We have been asked a number of times to present our results without using spinors, but in order to expose correctly the relations that occur in polarimetry we should have to invent a notation that at best would be no simpler than that of spinor algebra.  This paper shows spinor algebra at work in resolving the rather intricate interplay of misconceptions that came to represent the standard take on the BSA, which Hubbert \cite{hubbertreply} for example felt to be unacceptable.  There is however much more in polarimetry that spinors can help to resolve: typically oddities such as Huynen's pseudo-eigen problem \cite{huynen} and the \emph{missing} vector component in Cloude's theory for reciprocal target vectors \cite{bebbington:eusar} are things that have arisen because of imprecise associations that we hope to present as part of a much wider and coherent account in future papers on Geometric Polarimetry.

The organization of the paper is as follows.  In section
\ref{relations} we briefly outline some important aspects
regarding the origins of radar polarization algebra, and highlight
differences from Jones' optical polarization algebra. In
section \ref{critique} we comment and compare three significant formal approaches
to unitary transformations of polarization basis: the congruential transformation introduced by Graves \cite{graves}, the similarity transformation by Kostinski and Boerner \cite{kostinski}, and the consimilarity transformation proposed by L\"{u}neburg \cite{luneburg}. In section \ref{Graves} we demonstrate how the accepted congruential rule
can be justified in terms of the application of both mathematical
and physical principles without the need for antilinear
transformations. Finally, in section \ref{Examples} we present concrete examples that compare the congruential transformation with consimilarity and we show the wave reversal using spinors.

%% file: GP0Relations.tex

Radar polarimetry started to take off in the early 1950s, the post
World War II era when radar technology had become established, and
not long after the emergence of Jones's calculus for coherent
propagation in optical systems \cite{jonesI}. The states of plane
harmonic electromagnetic waves are described conventionally by
two-component complex vectors, now known as Jones vectors. They are complexified vectors, transverse to the wave-vector.
Ostensibly, we are dealing with Cartesian vectors that are
projected onto the phase-front, and we keep only the transverse
components. Geometrically, Jones vectors are considered to be Euclidean,
which implies several underlying assumptions, such as how to take
scalar products, and what defines orthogonality.

Throughout our discussions, unless otherwise stated, we will be
concerned with the description of complex harmonic plane-wave
fields and voltages with signature $\mbox{e}^{j(\omega
t-\mathbf{k}\cdot\mathbf{r})}$ for waves propagating in the
positive $\mathbf{r}$-direction.

In monostatic radar, one crucial difference with respect to the
majority of optical systems is that one must consider waves that
counterpropagate along the 'optical axis' of the system. To that end
Sinclair \cite{sinclair} introduced $2\times 2$ matrices, the scattering matrices, that
describe the far-field vector scattering of the wave by the target. They
are distinguished from the conventional Jones matrices
which instead describe transmission along a unidirectional optical axis. A
second even more crucial difference between the treatments of
optical and radar polarization that has received almost no formal
attention is that the latter needs to describe antenna as well as
field polarizations. The IEEE standard on polarization defines the
polarization of an antenna as that of the radiation that it
transmits, and also that to which it is maximally receptive
\cite{ieee83}. The fact that this definition is a statement of
labeling, but does not of itself prescribe a mathematical
relationship seems largely to have been overlooked.

From the outset, and even today, radar systems have been
predominantly monostatic, so that the problem most often dealt
with has been that of backscatter. Radar scientists for the most
part have preferred to adopt the backscatter alignment convention
(BSA), in which field components are measured in the same
transverse plane.

One of the complicating factors present in describing radar
scattering is that the field is detected indirectly - by measuring
voltage received by an antenna. Two linearly independent antennas
suffice to infer the field (in practice, it is usual to employ,
so-called orthogonal basis), such that the Jones vector components
representing the field in that basis provided are directly
proportional to the two complex voltages determined by an orthogonal
pair. No such analogue appears in the domain of Jones vectors in
optics, but it has generally been assumed that the concept of
antenna height \cite{mottI} can be complexified, and treated in the
same way as a Jones vector that describes the field. In the ensuing
critique of the aforementioned important papers regarding this, and
the transformation rules for various polarimetric quantities, it
will be seen that this assumption has not always been properly
questioned.

Antenna height is an extensive variable which relates the field
radiated by an antenna to the vectorial superposition of equivalent
dipole current vectors in the aperture plane.  A fundamental and
basic derivation shows that in a spherical coordinate system
$(\hat{\mathbf{r}},\hat{\mathbf{\theta}},\hat{\mathbf{\varphi}})$ in
which the antenna phase centre is at the origin, the radiated
far-field electric vector $\mathbf{E}$ is proportional to the
complex height vector $\mathbf{h}$\footnote{It is also called
effective length. It does not necessarily correspond to the physical
length of the antenna although there is a correspondence for a
dipole.} \cite{mottI}:
\begin{equation}\label{Ertf}
    \mathbf{E}\,(r,\theta,\varphi)=\frac{j Z_0 I}{2\lambda r}
    \,\mbox{e}^{-jkr}\,\mathbf{h}(\theta,\varphi),
\end{equation}
where $Z_0$ is the impedance of free space, $I$ the excitation
current, $\lambda$ the wavelength, $r$ is the distance from the
antenna ($\,r>>\lambda\,$), and the vectors are parallel to the
tangent plane at the observation point. What is significant here is
that the vector quantities $\mathbf{E}$ and $\mathbf{h}$ have identical direction. It is also
assumed that the ratio of the vector components may be complex.
Another relation considered to be fundamental in radar polarimetry
is that for a scattered field,
$\mathbf{E}=E_1\mathbf{e}_1+E_2\mathbf{e_2}$ received by an antenna
with complex height, $\mathbf{h}=h_1\mathbf{e}_1+h_2\mathbf{e}_2$,
measured in the same coordinates $\{\mathbf{e}_1,\mathbf{e}_2\}$, the received voltage $V$ is given
by
\begin{equation}\label{voltage}
    V=\mathbf{E}\cdot \mathbf{h}=E_1\,h_1+E_2\,h_2.
\end{equation}
Note that this is the \emph{ordinary} scalar product and not the Hermitian one which frequently arises in the theory of unitary invariants.
These relations are rigorously demonstrated in \cite{mottI}, who
uses standard Cartesian vector notation throughout. Complications
arise, however, when one wishes to consider unitary changes of
basis. Although one is never forced to change a basis, it is often
regarded as illuminating to do so, perhaps because of symmetries in
certain scattering problems. From a mathematical point of view, if
we know that certain representations should be unitarily equivalent,
we would also wish at least to be able to transform everything
consistently from one basis to another. When mathematicians or
physicists consider transformations, they usually demand and expect
that any relation in the theory should transform \emph{covariantly}.
Ultimately, this means that if everything in a relation is described
algebraically, then the algebra connecting all the transformed
objects should be identical to that of the original elements.  More
generally, we should expect to apply the same rule for any given
type of object.

%% file: GP0Critique.tex

We now look at three significant approaches from the radar
polarimetry literature to the description of unitary basis
transformation, in which we examine critically the assumptions on
which they are based. In the interests of ease of comparison we
have made slight notational changes regarding incident and
scattered states and such that inverse of the basis transformation
is applied in each case to the received electric field. These
changes make no substantive difference to the end results but
facilitate intercomparison.

\subsection{Congruential transformation}

In the radar polarimetry literature it was found that use of
conventional algebra to describe basis transformation seemed not to
be possible, and that new rules had to be invoked. Graves
\cite{graves} found it necessary to express outgoing and incoming
states under separate representations. He made use of \emph{direction vectors} $\mathbf{E}_+=(E_1,E_2)_+$ to describe a wave propagating in the positive $z$ direction and $\mathbf{E}_-=(E_1,E_2)_-$ in the negative $z$ direction where the components are referred to the same fixed reference basis. If the outgoing wave
$\mathbf{E}_+$ transforms by unitary basis change $Q$ as in the new
outgoing wave $\mathbf{E}_{+}'$,
\begin{equation}\label{}
    \mathbf{E}_{+}'=Q^{-1}\mathbf{E}_+
\end{equation}
then, according to Graves, incoming wave $\mathbf{E}_-$ transforms
as
\begin{equation}\label{}
    \mathbf{E}_{-}'=\bar{Q}^{-1}\mathbf{E}_-.
\end{equation}   
where we use overbar (rather than the more usual asterisk) to
denote complex conjugation - this notation, more often seen in the mathematical literature, is convenient when we want to be clear when an entire expression is conjugated.

Graves then goes on to give the widely accepted congruential rule for transformation of the backscatter matrix,
\begin{equation}\label{Gravesrule}
    S'=USU^T
\end{equation}
where, for the sake of uniformity of treatment, we denote the basis
transformation matrix as $U=Q^T$. From here on, we shall refer to (\ref{Gravesrule}) as \emph{Graves' rule}. In his argument, Graves uses the equivalence $\bar{U}^{-1}=U^T$ for the last step, as far as we
can tell for no other reason than that $U^T$ is neater. Graves does not employ the concept of antenna
height, but argues that the received voltage is expressed as the
(non-Hermitian) scalar product of the normalised $\mathbf{E}_+$
corresponding to the transmitted wave and $\mathbf{E}_-$
representation of the scattered wave. No physical or mathematical
justification is provided as to what a scalar product between two
fields might actually mean. The relation (\ref{Gravesrule}) turns
out to be the generally accepted result, although as we remark, the
way in which it is arrived at lacks rigour in terms of justification
of assumptions. We defer further discussion to the last part of this
section where L\"{u}neburg's appeal to consimilarity makes the
assumptions explicit.

\subsection{Similarity transformation}\label{similarity:sec}

Much later, Kostinski and Boerner \cite{kostinski} considered the
rules of transformation for field vectors, antenna heights and
scattering matrices. The rules they applied were,
\begin{equation}\label{Erules}
    \begin{array}{l}
      \mathbf{E}_{\text{\tiny{R}}}^{'} = U\mathbf{E}_{\text{\tiny{R}}} \\
      \mathbf{E}_{\text{\tiny{T}}}^{'} = U\mathbf{E}_{\text{\tiny{T}}} \\
      \mathbf{h}^{'} \;\,= U\mathbf{h} \\
    \end{array}
\end{equation}
from which they deduced a transformation rule that is not a
congruential transformation, but a \emph{similarity}:
\begin{equation}\label{similarity}
    S^{'}=USU^{-1}.
\end{equation}
While acknowledging a second representation obtained by transforming
the voltage law, which yields the usual congruential transformation
(\ref{Gravesrule}), Kostinski and Boerner stated a preference for
the similarity transformation, even though, as they noted, the
transformation rule they deduced for the voltage equation
(\ref{voltage}) now apparently fails to transform covariantly:
\begin{equation}\label{voltagetransform}
    V=\mathbf{h}^{'T}\cdot U^T U \mathbf{E}_{\text{\tiny{R}}}^{'}.
\end{equation}

Mieras \cite{mieras} criticized this choice, preferring the
congruential rule. He argued that it would be preferable to work
with the voltage equation without reference to the received field.
Nevertheless, Mieras accepted the derivation of
(\ref{voltagetransform}) as algebraically valid. From a
mathematical point of view, however, an expression such as
(\ref{voltagetransform}) as the outcome of a basis change is
anathema. A scalar relation should be algebraically invariant
under transformation, because this is the whole rationale of basis change.

In all the arguments presented so far, there has been an implicit
assumption that (\ref{Ertf}) transforms alike on each side under
unitary basis change. By assuming it, Graves in fact made no
algebraic distinction between field vectors and antenna heights
and, as we noted, introduced without question or justification the
scalar product between two field states to obtain the voltage.
From (\ref{Erules}) it can be seen that essentially Kostinski and
Boerner made the same assumption. Given their starting point they
were not incorrect to state that the scattering matrix connecting
transmitted fields should transform as a similarity. In doing so
they were abandoning Graves' use of directional Jones vectors, and
applying globally the same transformational rules for all fields.
In effect, they were applying the principles of tensor algebra, in
which linear relationships are supposed to be expressible
regardless of basis. Working in a regime of Cartesian vector
analysis one normally has no need to make any kind of distinction
between different kinds of vector, the reason being that inner
products are invariant under rotation, (or other Euclidean
isometries). Actually, (\ref{voltagetransform}) does transform
satisfactorily if one restricts to rotations (for which $U^T=U^{-1}$), but not under
general unitary transformations. Lack of general unitary
covariance\footnote{Covariance of a law is the invariance of the form of the law under coordinate transformations.} in (\ref{voltagetransform}) stems from the assumption
that the last relation in (\ref{Erules}) is valid, namely that
antenna height vectors transform generally in the same way that
field vectors do. To question this, however, appears to throw into
doubt the apparently fundamental assumption of (\ref{Ertf}). In
fact there is a good theoretical reason why, although (\ref{Ertf})
is true when expressed in a linear basis, it does not generalise
under unitary transformation. This is because the fundamental
derivation of (\ref{Ertf}) arises by applying the free-space
dyadic Green's function $\overleftrightarrow{\mathbf{G}}$ to the
elementary current element:
\begin{eqnarray}
  \mathbf{E}(\mathbf{r})&=& \int\!\!\!\int\!\!\!\int\overleftrightarrow{\mathbf{G}}\cdot
    \mathbf{J}(\mathbf{r}')\,dr'^3= \nonumber \\
   &=& \int\!\!\!\int\!\!\!\int(\overleftrightarrow{\mathbf{I}}+k^{-2}\nabla\nabla)\,\frac{\mbox{e}^{-jkR}}{R}\,
    \mathbf{J}(\mathbf{r}')\,dr'^3, \label{EGreen}
\end{eqnarray}
where $R=|\,\mathbf{r}-\mathbf{r}^{'}|$, $k$ is the wavevector magnitude, $\mathbf{J}$ is the
current density which is the source of the field and $\overleftrightarrow{\mathbf{G}}$ is the dyadic homogeneous Green function given by
\begin{equation}\label{}
    \overleftrightarrow{\mathbf{G}}=\left(\overleftrightarrow{\mathbf{I}}+k^{-2}\nabla\nabla\right)\frac{\mbox{e}^{-jkR}}{R},
\end{equation}
$\frac{\mbox{e}^{-jkR}}{R}$ is the scalar Green function and $\overleftrightarrow{\mathbf{I}}$ is the unit dyadic.
The integration in (\ref{EGreen}) takes place over the entire antenna. In the spatial Fourier
domain, the dyadic factor in the Greens function takes the form,
\begin{equation}\label{dyadic}
    \overleftrightarrow{\mathbf{I}}-\frac{1}{k^2}\overleftarrow{\mathbf{k}}\overrightarrow{\mathbf{k}}
\end{equation}
which has the form of a projection operator that projects out the
longitudinal components of the current vector and ensures that the
far field is transverse. If the equivalent antenna height is already
presumed to be transverse, the second term in (\ref{dyadic}) may
apparently be omitted with impunity leaving only the unit dyadic,
which may, in a purely conventional derivation also be omitted.
However, the unit dyadic is not invariant under unitary
$\mathrm{SU}(2)$ congruential transformation. From a tensorial point
of view, the unit dyadic is really a covariant Euclidean metric
tensor\footnote{In fact dyads extend vectors to provide an alternative description to second rank tensors and the use of dyadics is nearly archaic since tensors perform the same function but are notationally simpler.},
\begin{equation}\label{}
    \overleftrightarrow{\mathbf{I}}=\overleftarrow{\mathbf{e}_x}\overrightarrow{\mathbf{e}_x}+\overleftarrow{\mathbf{e}_y}\overrightarrow{\mathbf{e}_y}+\overleftarrow{\mathbf{e}_z}\overrightarrow{\mathbf{e}_z}
\end{equation}
built with the basis and the reciprocal basis, and it is only invariant under Euclidean isometries. In tensor
language, we say that (in either time- or spatial- frequency domain)
the Green's function operator lowers the index of the vector it
operates on. Thus, if it operates effectively on the antenna height
vector (in tensor language an upper index contravariant vector), it
produces a lower index covariant vector which shows that the electric field has the
same tensor character as that of the gradient of a potential, which
it has in the static case. Tensor-invariant descriptions require
that contravariant and covariant objects transform contragrediently
- essentially by mutually inverse transformations. To express this
in matrix algebra requires in addition a transpose in one of the
matrices as matrix multiplication occurs always from the left. This
is the essential missing ingredient in the argument that erroneously
leads to (\ref{voltagetransform}).

As Mieras \cite{mieras} correctly noted, the covariant
transformation law of the voltage equation is fundamental. The widely
accepted form of transformation law for the backscatter matrix as
a congruential relation can be properly explained by defining the
backscatter matrix via the relation (see \cite{mottI}), (absorbing the range factor into the scattering matrix),
\begin{equation}\label{voltagehh}
    V(\mathbf{h}_{\text{\tiny{R}}},\mathbf{h}_{\text{\tiny{T}}})=\mathbf{h}_{\text{\tiny{R}}}\cdot S\,\mathbf{h}_{\text{\tiny{T}}}=\mathbf{h}_{\text{\tiny{T}}}\cdot
    S\,\mathbf{h}_{\text{\tiny{R}}}
\end{equation}
for unit current excitation. The symmetry of $S$ is obviously
implied by reciprocity, and the congruential transformation rule
is consequently automatic. Understanding $S$ as a bilinear form is
conceptually quite different from the usual meaning of $S$ as a
linear operator on the incident field; indeed, it would probably
be more proper to call the operator in (\ref{voltagehh}) something
other than the scattering matrix, even if, componentwise it is
equal to it. We should also note that application of
(\ref{voltagehh}) is not restricted to backscatter; then, of
course, the matrix is no longer symmetric (since the antennas are
not co-located) although reciprocity is reflected in the
interchangeability of transmission and reception when the value of
the form vanishes \cite{bebbington:invariance}.

What has not yet been accomplished, however, is an explanation of
how Graves' rule applies to the scattering matrix proper, and we
address this in section \ref{Graves}.

\subsection{Consimilarity transformation}

At this point, it is appropriate to consider the proposal of
L\"{u}neburg that the BSA form of the backscatter equation should
be considered to transform as a consimilarity relation
\cite{luneburg:revision}. We need seriously to examine whether
Graves' rule for scattering matrices proper turns out indeed to be
a special case of consimilarity. The mathematics of consimilarity
is described in Horn and Johnson \cite{hornjohnson}. In arguing
for this as a principle, one is saying in effect that the
fundamental rule for transforming the scattering matrix is the
consimilarity,
\begin{equation}\label{Lune}
    S^{'}=AS\bar{A}^{-1}.
\end{equation}
This relation implies (as did Graves) that the counterpropagating
waves transform conjugately with respect to each another.
L\"{u}neburg's argument is that wave reversal can be made
equivalent to time reversal (justified by the symmetry of
Maxwell's equations in vacuo under time reversal) via the relation
\cite{luneburg:finalreport}
\begin{equation}\label{ReE}
    \mbox{Re}\,[\mathbf{E}\:\mbox{e}^{j(\omega
    t-kz)}]\;\stackrel{T:\,t\rightarrow -t}{\longrightarrow}\; \mbox{Re}\,[\mathbf{E}\:\mbox{e}^{-j(\omega
    t+kz)}]= \mbox{Re}\,[\bar{\mathbf{E}}\:\mbox{e}^{j(\omega
    t+kz)}],
\end{equation}
where T stands for the time-reversal operator. The stated position
here is that conjugation effects wave reversal without change in
the polarization label. L\"{u}neburg's equations therefore
explicitly include an antilinear operation (T: time reversal, or
conjugation), which implies \cite{luneburg:finalreport},
\begin{equation}\label{antilinear}
    T\,(\alpha \mathbf{u}+\beta
    \mathbf{v})=\bar{\alpha}\,T(\mathbf{u})+\bar{\beta}
    \,T(\mathbf{v}).
\end{equation}
In the end L\"{u}neburg notes that, for unitary transformations $A\,\rightarrow \, U\,\,\, / \,\,\, \bar{U}^{T}=U^{-1}$, we have that
\begin{equation}\label{}
    \bar{A}^{-1} \, \rightarrow U^T
\end{equation}
so arriving at Graves' rule (\ref{Lune}). L\"{u}neburg therefore effectively
formalises Graves' treatment, by clarifying that an antilinear
operation is required as a formal operator in the scattering
equation, but reverts to the congruential form of Graves' rule in
the unitary case. Physically, however, arguments in favour of an
antilinear operation are not acceptable in the sense that
(\ref{similarity}) is regarded as a special case of a
consimilarity. The problem with this is that time-reversal is only
a symmetry of Maxwell's equations in vacuo or at any rate in
certain lossless media, in other words when unitarity is
guaranteed. In real media, however, such as the atmosphere, the
presence of an anisotropic lossy medium (e.g. when precipitation is
present on the path) is an important factor to consider; for
weather radars in particular, an unignorable factor. There it has long been established rigorously \cite{mccormick:priciples}, \cite{mccormick:prop} that
Graves' rule is a special case of the active transformation,
\begin{equation}\label{}
    S'=ASA^T
\end{equation}
where $A$ is the propagation matrix between target and radar
acting on the received field. This holds for a medium exhibiting
reciprocity, includes the case of multiple forward scatter
(coherent propagation). At S-band frequencies, it was shown
\cite{bebbington:correction} that the effective scattering matrix
transformation induced by forward propagation through rain is very
close to a unitary one of this form. Approximate time-reversal
symmetry appropriately describes \emph{phase conjugation}, a
technique that is realised in modern optics to reverse a wavefront
so that it substantially retraces its path. This requires
non-linear processes (a non-linear medium, together with optical
pump-waves). In such a medium, exhibiting for example the
high-frequency Kerr effect \cite{akhmanov}, the dielectric tensor
describes interactions between the incident wave and pump waves,
such that one must use the full representation of the signal, not
just the analytic representation. In essence, then, frequency
conversion may occur through non-linear processes such that the
component of the real signal associated with the conjugate phase
propagates with the reversed wave vector. Nieto-Vesperinas
\cite{nieto} uses in effect the same device as L\"{u}neburg
(\ref{ReE}) to describe the conjugated wave. It must be clear,
however, that this can only occur through non-linear interactions
which are not a part of normal radar scattering theory. Moreover,
the usual realizations of phase-conjugate mirrors employ
degenerate (or near-degenerate) four-wave mixing (DWFM), in which
the incident wave is scattered by a time-dependent Bragg grating
oscillating at twice the nominal signal frequency. Thus, for an
incident wave at frequency $\omega_i$ and a pump signal at
$\omega_p\thickapprox \omega$, the scattered signal is at
$\omega_s=2\omega_p-\omega_i$ resulting in sideband reversal,
which is not a feature of linear scattering. A phase conjugated
wavefront also propagates differently from a normal one in an
inhomogeneous medium, so from almost all perspectives,
consimilarity applied globally does not describe the physics. In
addition, equation (\ref{antilinear}) does not respect the
expected linear superposition rule. We can hardly emphasise
sufficiently strongly that, in dealing with linear systems, the
introduction, whether explicitly or implicity (e.g. via partial
conjugation of a term) of antilinearity lacks any physical justification.
It may be noted that we have restricted attention so far (apart
from in the preceding example) to media exhibiting reciprocity. In
discussing generalizations we must be aware that there are partial
and fuller generalizations concerning the effects of propagation.
In the context of remote sensing it is reasonable also to consider
effects of non-reciprocity arising from propagation through the
ionosphere, i.e. Faraday rotation. As it turns out, although
Faraday rotation is a unitary process, a medium exhibiting this
phenomenon does not have time reversible symmetry, as the
imaginary components of the Hermitian constitutive tensor reverse
\cite{nieto}. The implications for the congruential rule, on the
other hand, are that whilst in a reciprocal medium the propagation
matrix is insensitive to direction, the sign of anisotropy induced
by magnetic Faraday rotation depends (in any coordinate system) on
the projection of the wave vector along magnetic field. For
example, in a linear basis, if the Faraday rotation is represented
by a rotation matrix $R(\alpha)$ where $\alpha$ is the rotation
measure, then,
\begin{equation}\label{}
    S^{'}=R(\alpha)\,S\,R^T(-\alpha),
\end{equation}
while in a circular basis, we have,
\begin{equation}\label{}
    S^{'}=\begin{pmatrix}
      \mbox{e}^{j\alpha} & 0 \\
      0 & \mbox{e}^{-j\alpha} \\
    \end{pmatrix}\,S\,\begin{pmatrix}
      \mbox{e}^{j\alpha} & 0 \\
      0 & \mbox{e}^{-j\alpha} \\
    \end{pmatrix}^T.
\end{equation}
In each case, the effect of rotation measure accumulates over both
outward and return paths. This is no longer a congruential rule,
but is a straightforward generalization similar to the situation
in bistatic scattering, where the propagation factors on each path
generally differ. As usual, the presence of the transpose is
nothing more than an artefact of left-right ordering in matrix
algebra.

%% file: GP0Graves.tex

Having dismissed the premises on which Graves' rule was
traditionally based, we now turn to providing an alternative
explanation that requires no suggestion of antilinearity. That is,
we explain, how the relation
\begin{equation}\label{}
    \mathbf{E}^{'}_{\text{\tiny{R}}}=S\mathbf{E}_{\text{\tiny{T}}}
\end{equation}
in matrix algebra comes to have a congruential transformation
rule. Now that we have explained how, fundamentally, we should
come to regard the field vectors as covariant (see section \ref{similarity:sec}), the congruential
rule for transforming $S$ appears strange. There is still a
requirement to introduce some form of directional Jones vector,
such that the fields have different representations, however this
cannot match the requirements of the BSA convention by means of a
simple spatial rotation.

An example provided in Mott's text \cite{mottI} is very
instructive in this. Although Mott does not give a general rule
for basis transformation, he describes a relation for transforming
from a linear to a circular basis \cite{mottI} (p236) as,
\begin{equation}\label{}
    \begin{pmatrix}
      E_x-jE_y \\
      E_x+jE_y \\
    \end{pmatrix}_R=\begin{pmatrix}
      A_{RR} & A_{RL} \\
      A_{LR} & A_{LL} \\
    \end{pmatrix}\begin{pmatrix}
      E_x+jE_y \\
      E_x-jE_y \\
    \end{pmatrix}_T,
\end{equation}
by substituting on the left hand side in terms of the original
matrix elements, so obtaining a symmetric matrix in terms of them.
Because this is not expressed explicitly as a matrix
transformation, it is easy to see in this instance that the end
result is obtained by assuming from the outset that the Cartesian
representations for left and right handed polarizations are
exchanged on reversal. It should be particularly noted that there
is no explicit suggestion of conjugation here, it is purely a
question of representation.

Now we move on to address the more general case of arbitrary basis
change. To do this, it is necessary to introduce some aspects of
spinor analysis, which we intend to bring into play much more
extensively in our new approach of Geometric Polarization. This, we
believe provides a very satisfactory explanation for the change in
representation from a geometric perspective and which was suggested as an alternative to tensor representation in this context in \cite{alvarez}. To illustrate how
such methods can justify the algebraic procedures, we need only to introduce some of the most basic spinorial concepts that help to
show how the geometry of space can be described in terms of
objects even more basic than ordinary vectors. The origins of
spinor techniques in physics goes back to the origins of
relativistic quantum mechanics as new mathematical methods were
required to explain quantum mechanical spin, in particular the
fact that the wave function of a 'spin-$\frac{1}{2}$' fermionic
particle changes sign on rotation by $180^{\circ}$, a fact that
cannot be explained by conventional vector and tensor analysis
alone \cite{post}.

Spinors were introduced by Cartan \cite{cartan} in a geometric context. They are not vectors, but vectors can be constructed from products of spinors. Spinors can therefore be seen as in some sense more fundamental.  From the polarimetric point of view it is important to consider that ultimately both the fields in which we are interested and the reference frame in which their components are expressed can be related unambiguously to a common spinorial reference system, a spin-frame, comprising an ordered pair of spinors.  Polarimetry is as much about geometry as it is about electromagnetics, and it is a major strength of the spinor approach that it it gives such a tight relationship to the spatial geometry of the frame, not just to the plane of polarization.

Woodhouse \cite{woodhouse} provides a very succinct introduction
to spinors, Payne \cite{payne} describes spinors in an elementary way using trigonometry, while Penrose and Rindler \cite{penrose} provides a
very comprehensive description and is generally regarded as a
standard reference. We adopt the same notation as these authors,
as it now appears to be a \emph{de facto} standard in the physical
literature. Our usage differs in one respect alone, owing to the
fact that our analytical signal signature which conforms with the
majority usage in the engineering literature is the opposite of
that generally adopted in the theoretical physics literature.
Since they also use the symbol '$i$' for the imaginary unit, we
can consistently translate everything by replacing our imaginary
'$j$' for '$-i$'. The connections between spinors and space-time
$4-$vectors mirror in a way that we can regard as non-accidental
that between Jones vectors and Stokes vectors, so the basic
geometric concepts are in principle easy to assimilate from a
knowledge of polarimetry.

Often space-time is considered to be spanned by unit vectors,
\begin{equation}\label{}
    \hat{\mathbf{t}}=\begin{pmatrix}
      1 \\
      0 \\
      0 \\
      0 \\
    \end{pmatrix}, \quad \hat{\mathbf{x}}=\begin{pmatrix}
      0 \\
      1 \\
      0 \\
      0 \\
    \end{pmatrix}, \quad\hat{\mathbf{y}}=\begin{pmatrix}
      0 \\
      0 \\
      1 \\
      0\\
    \end{pmatrix}, \quad\hat{\mathbf{z}}=\begin{pmatrix}
      0 \\
      0 \\
      0 \\
      1 \\
    \end{pmatrix},
\end{equation}
along the time and orthogonal Cartesian spatial axes. Matters are
greatly simplified if time and space are measured in the same
units, which is rather what radar engineers do, in converting time
into distance, and physicists' language for the same process is to
say we choose units such that the speed of light is unity.

An alternative, but admissible basis, frequently adopted in the
spinor literature, uses instead four null-vectors,
\begin{equation}\label{lnm}
    \begin{array}{l}
      \mathbf{l}\,\,\,=\hat{\mathbf{t}}-\hat{\mathbf{z}} \\
      \mathbf{n}\,=\hat{\mathbf{t}}+\hat{\mathbf{z}} \\
      \mathbf{m}=\hat{\mathbf{x}}-j\hat{\mathbf{y}} \\
      \bar{\mathbf{m}}=\hat{\mathbf{x}}+j\hat{\mathbf{y}} \\
    \end{array}.
\end{equation}
They are self-orthogonal, but not all mutually orthogonal,
although $\mathbf{l}$ and $\mathbf{n}$ are orthogonal to
$\mathbf{m}$ and $\bar{\mathbf{m}}$.

\begin{figure}[h]
                        \centering
                        \includegraphics[width=1.5in]{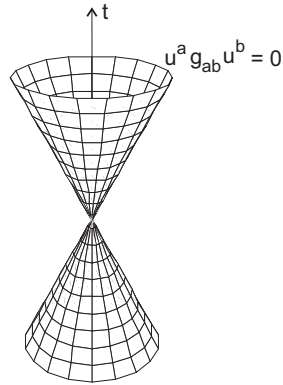}
                        \caption{The light-cone.}
                        \label{cono}
\end{figure}

The Lorentz metric tensor of special relativity,
\begin{equation}\label{gab}
    g_{ab}=\begin{pmatrix}
      1 & 0 & 0 & 0 \\
      0 & -1 & 0 & 0 \\
      0 & 0 & -1 & 0 \\
      0 & 0 & 0 & -1 \\
    \end{pmatrix}
\end{equation}
is required to describe the squared distance or interval between
space-time points as,
\begin{eqnarray}
  ds^2 &=& dt^2-dx^2-dy^2-dz^2= \nonumber \\
   &=& \begin{pmatrix}
      dt \\
      dx \\
      dy \\
      dz \\
    \end{pmatrix}^T\begin{pmatrix}
      1 & 0 & 0 & 0 \\
      0 & -1 & 0 & 0 \\
      0 & 0 & -1 & 0 \\
      0 & 0 & 0 & -1 \\
    \end{pmatrix}\begin{pmatrix}
      dt \\
      dx \\
      dy \\
      dz \\
    \end{pmatrix} \label{ds2}
\end{eqnarray}
which is a symmetric quadratic form. By convention, timelike
intervals are positive, while spacelike intervals are negative
(note that many, especially older, texts apply the reverse sign
convention). The vanishing intervals are referred to as lightlike
as this condition effectively says that two points so separated
mark the start and endpoint of a path segment of a light signal.
By virtue of (\ref{ds2}) we have, the analytical verification of
the nullity of the basis vectors in (\ref{lnm}),
\begin{equation}\label{}
    u^ag_{ab}u^b=0, \quad \text{where} \quad u^ag_{ab}u^b\equiv \sum_{a=0}^3 \,\sum_{b=0}^3 u^ag_{ab}u^b
\end{equation}
when $u^a$ describes in tensor index form any of the null basis
vectors. Here we are adopting as standard the Einstein summation
convention where a summation sign over index values $0-3$ is
implied where there are repeated indices, one in the upper
position (denoting a contravariant index) and one in the lower
position, (denoting a covariant index).
All these vectors lie on
the so-called light-cone (all light-ray trajectories through the
spacetime origin \cite{penrose}), which is a double cone whose
vertex is the origin which separates its past and future parts as shown in Fig. \ref{cono}.
This language can be helpfully applied to the problem of
scattering, as we might say that the signal incident on a
scatterer propagates on its past light-cone, while the scattered
wave lies on its future light-cone. By extension any multiples of a
null-basis vector also lies on the null-cone although, generally
speaking, arbitrary linear combinations of them do not. Real
null-vectors are also said to be isotropic \cite{post}, and Cartan \cite{cartan}
effectively developed the idea of spinors in three dimensions as
representations of such vectors by ordered pairs of complex
numbers.

The reason for adopting a null-tetrad as a basis is that in a
spinorial description these have a particularly simple
representation. Spinors are simply complex $2-$vectors that span
spin-space, their first application in physics being by Pauli
\cite{pauli} in terms of spin matrices, also well-known in modern
polarimetric theory \cite{cloude:liegroups:journal}.
Conventionally, the basis \emph{vectors} of this space are denoted,
\begin{equation}\label{contraspinbasis}
    o^A\equiv \begin{pmatrix}
      o^0 \\
      o^1 \\
    \end{pmatrix}= \begin{pmatrix}
      1 \\
      0 \\
    \end{pmatrix}, \quad \iota^A\equiv \begin{pmatrix}
      \iota^0 \\
      \iota^1 \\
    \end{pmatrix}=\begin{pmatrix}
      0 \\
      1 \\
    \end{pmatrix}.
\end{equation}
Here the uppercase index $^A$ conventionally takes values $0$ and
1. Furthermore, we require the conjugates of these (which are
numerically equal, but must be treated as belonging to a distinct
conjugate representation of spin space),
\begin{equation}\label{}
    o^{A'}\equiv \begin{pmatrix}
      1 \\
      0 \\
    \end{pmatrix}, \quad \iota^{A'}\equiv \begin{pmatrix}
      0 \\
      1 \\
    \end{pmatrix}.
\end{equation}

The use of a prime $^{'}$ on the index is a book-keeping device
to ensure that objects so labelled are transformed conjugately
with respect to unprimed indices.

The spinor representations of the null-vectors can be taken as
\cite{penrose},
\begin{equation}\label{lnm:spinor}
    \begin{array}{l}
      \mathbf{l} \;\,\,\, \rightarrow \; o^A o^{A'}\equiv \begin{pmatrix}
        o^0 o^{0'} & o^0 o^{1'} \\
        o^1 o^{0'} & o^1 o^{1'} \\
      \end{pmatrix}=\begin{pmatrix}
        1 & 0 \\
        0 & 0 \\
      \end{pmatrix}\\
       \mathbf{n} \; \,\rightarrow \; \iota^A \iota^{A'}\,\equiv \begin{pmatrix}
        0 & 0 \\
        0 & 1 \\
      \end{pmatrix} \\
       \mathbf{m} \; \rightarrow \; o^A \iota^{A'}\equiv \begin{pmatrix}
        0 & 1 \\
        0 & 0 \\
      \end{pmatrix} \\
       \bar{\mathbf{m}} \; \rightarrow \; \iota^A o^{A'}\equiv \begin{pmatrix}
        0 & 0 \\
        1 & 0 \\
      \end{pmatrix} \\
    \end{array}.
\end{equation}
These are rank-one (singular) matrices, their singularity being
equivalent to the nullity of the corresponding vectors.

To complete the material necessary to our analysis, we also need
to mention the 'metric' spinor, which unlike that of space-time is
skew. The metric spinor, in both contravariant and covariant
forms, is given by,
\begin{equation}\label{}
    \varepsilon^{AB}\equiv \begin{pmatrix}
      \quad\! 0 & 1 \\
      -1 & 0 \\
    \end{pmatrix}, \quad  \varepsilon_{AB}\equiv \begin{pmatrix}
      \quad\! 0 & 1 \\
      -1 & 0 \\
    \end{pmatrix}
\end{equation}
and can be viewed either as a computational device to make an inner
product between pairs of covariant spinors or pairs of
contravariant spinors, or as a means of converting a spinor of one
type into one of the other (in which case, one often retains the
identifying symbol to show it as a mapping or equivalence
relation. Thus,
\begin{equation}\label{covariantspinbasis}
    o_B=o^A \varepsilon_{AB}\equiv \sum_{A=0}^1 o^A \varepsilon_{AB}=\begin{pmatrix}
      0 \\
      1 \\
    \end{pmatrix}, \quad \iota_B=\iota^A \varepsilon_{AB}=\begin{pmatrix}
      -1 \\
      \quad \!0 \\
    \end{pmatrix}
\end{equation}
represent covariant counterparts to the contravariant basis
elements. Here the summation convention operates in the same way
as with tensors, but over index values $0-1$.

We consider now the so-called priming
operation discussed in Penrose and Rindler \cite{penrose} (p262) in which certain spinor operators may be
expressed by means of a re-assignment of the spinor indices. Here
we adopt a slightly different operation, and the primed basis spinors $o^{\tilde{A}}$, $\iota^{\tilde{A}}$ are defined as:
\begin{equation}\label{otilde}
    o^{\tilde{A}}\;\rightarrow \;-\iota^A, \quad \iota^{\tilde{A}}\;\rightarrow \;o^A
\end{equation}
(instead of $jo^{\tilde{A}}\;\rightarrow \;\iota^A$,
$j\iota^{\tilde{A}}\;\rightarrow \;o^A$), which has most of the same
properties including preservation of the orientation of the
spin-frame. Under this operation, the following transformation of
the null-basis vectors takes place (see Appendix):
\begin{equation}\label{lnmpriming}
    \begin{array}{l}
      \tilde{\mathbf{l}}\;\,\,\,\rightarrow\;\mathbf{n} \\
       \tilde{\mathbf{n}}\:\,\,\rightarrow\;\mathbf{l} \\
       \tilde{\mathbf{m}}\;\rightarrow\;-\bar{\mathbf{m}} \\
       \tilde{\bar{\mathbf{m}}}\;\rightarrow\;-\mathbf{m} \\
    \end{array}.
\end{equation}
The interpretation of this is that the past and future null-cones
are interchanged, and (apart from an absorbable sign-change) left
and right handed null-vectors that are orthogonal to $\mathbf{l}$
and $\mathbf{n}$ are also interchanged. Overall, because of two
interchanges there is no inversion of the orientation of
space-time, and antilinear spinor transformations are not
required.

While interpreting our basis vectors as a frame suitable for
describing waves traveling along the $z-$axis, $\mathbf{l}$ and
$\mathbf{n}$ describe the space-time directions of constant phase
trajectories of outgoing and incoming plane waves, while
$\mathbf{m}$ and $\bar{\mathbf{m}}$ form a complex basis for
planes parallel to the wavefront. That interchange of $\mathbf{m}$
and $\bar{\mathbf{m}}$ reverses the orientation of such a plane is
evident, as (use of conventional Cartesian vector calculus is
justified by the fact that $\mathbf{m}$ and $\bar{\mathbf{m}}$ are
purely spatial),
\begin{equation}\label{}
    \begin{array}{l}
      \bar{\mathbf{m}}\times\mathbf{m}=(\hat{\mathbf{x}}+j\hat{\mathbf{y}})\times(\hat{\mathbf{x}}-j\hat{\mathbf{y}})=-2j\,\hat{\mathbf{x}}\times \hat{\mathbf{y}} \\
      \mathbf{m}\times\bar{\mathbf{m}}=(\hat{\mathbf{x}}-j\hat{\mathbf{y}})\times(\hat{\mathbf{x}}+j\hat{\mathbf{y}})=+2j\,\hat{\mathbf{x}}\times \hat{\mathbf{y}}
      \\
    \end{array}.
\end{equation}

The two complex basis vectors $\mathbf{m}$ and $\bar{\mathbf{m}}$
are eigenvectors of matrices describing plane rotations and so we
can interpret the effect of the priming operation as exchanging
the sense of rotation or angle measure in the plane from the
perspective of a fixed orientation.

The final step required in the argument to explain Graves' rule,
is to note the effect of the priming operation (\ref{otilde}) on
matrix representations.

In spinor algebra, matrix operations on covariant spinors take the
form \cite{woodhouse}
\begin{equation}\label{eta}
    \eta_A=M_A^{\,\,\,\,B}\xi_B.
\end{equation}
Such matrices map covariant spinors to covariant spinors. We can
form a basis for all such matrices from outer products of the
basis spinors, the first of each pair in covariant form, and the
second contravariant. A basis is therefore obtained
using the set,
\begin{equation}\label{}
    o_Ao^B,\quad o_A\iota^B, \quad \iota_Ao^B, \quad
    \iota_A\iota^B.
\end{equation}
Note that the conjugated bases do not appear in these expressions.
Note, also that these are matrices not scalars because summation
is not implied due to the index labels being different (summation
would give the matrix trace). Under the priming operation we form
an equivalent basis,
\begin{equation}\label{}
    o_{\tilde{A}}o^{\tilde{B}},\quad o_{\tilde{A}}\iota^{\tilde{B}}, \quad \iota_{\tilde{A}}o^{\tilde{B}}, \quad
    \iota_{\tilde{A}}\iota^{\tilde{B}}.
\end{equation}
Via this equivalence relation, a general matrix transforms as (the
algebra of expressions (\ref{eta}) - (\ref{alpha}) is expanded in
full in Appendix A.)
\begin{equation}\label{alpha}
    \begin{pmatrix}
      \alpha & \beta \\
      \gamma & \delta \\
    \end{pmatrix} \quad \rightarrow \quad \begin{pmatrix}
      \,\,\,\delta & -\gamma \\
      -\beta & \,\,\,\alpha \\
    \end{pmatrix}.
\end{equation}
This operation described in matrix language is the transpose of
the adjugate. Note that it is a completely linear mapping. If we
restrict attention to unimodular, or a fortiori, unitary matrices,
this becomes the mapping\footnote{In fact the adjugate is used to compute the determinant of a matrix: $\det (M)=M \;\mbox{adj}(M)$. For unimodular matrices, $\mbox{adj}(M)=M^{-1}$.},
\begin{equation}\label{}
    U \;\rightarrow \; (U^{-1})^T.
\end{equation}
Thus, it can be seen that under this mapping, if $U$ describes the
unitary matrix that effects a basis change on the received field
vector, then for the voltage equation to be invariant, the antenna
height vector must transform via the inverse transpose. But if the
priming operation describes the change of reference for Graves'
outgoing directional Jones vector, we see that the representation
of the transmit field transforms in the same way as (receive or
transmit) antenna height vector. Then, since
\begin{equation}\label{Vo}
    V=\mathbf{h}_{\text{\tiny{R}}}^T\, S \, \mathbf{h}_{\text{\tiny{T}}} \quad \Rightarrow \quad
    \mathbf{E}_{\text{\tiny{R}}}=S\,\mathbf{h}_{\text{\tiny{T}}},
\end{equation}
then identification of the components of $\mathbf{E}_{\text{\tiny{T}}}$ and
$\mathbf{h}$ in any one basis implies that they are equal in any.

To summarize, the priming operation, as a linear mapping, allows
(\ref{Ertf}) and (\ref{voltage}) to be simultaneously true. The
story is rather more complicated than if we had simply described
the scattering process as a bilinear form, but we can express
concepts of basis invariance without violation of either physical
or mathematical principles.

%% file: GP0Examples.tex

\subsection{Wave reversal}

First, we demonstrate how the spin-frame manipulations can be handled for representing the antenna height, and for the propagating fields in reception and transmission. The case of circular polarizations is among those most likely to engender confusion, so we work this through in detail.  Consider as our primary object associated with a left-hand circular polarization, the antenna state, for which we construct the antenna height spinor representation $\eta^A$ in an (H=0,V=1) basis,

\begin{equation}\label{}
    \eta^A=\begin{pmatrix}
      \eta^0 \\
      \eta^1 \\
    \end{pmatrix}=\frac{1}{\sqrt{2}}\begin{pmatrix}
      1 \\
      j \\
    \end{pmatrix}=\frac{1}{\sqrt{2}}\left(o^A+j\iota^A\right),
\end{equation}
where the basis spinor $o^A$ and $i^A$ are specified in (\ref{contraspinbasis}).
By convention, the polarization state that the antenna is most receptive to is defined to be the same as that which it transmits.  It is, in any case empirically verifiable that circular polarization antennas receive waves of the same handedness that they transmit.  In order for the scalar voltage received at the antenna to be invariant under unitary change of basis, it is therefore clear that the spinor representation for the unit covariant LHC electric field state incident on the antenna must be,
\begin{equation}\label{}
    \psi_A=\begin{pmatrix}
      \psi_0 \\
      \psi_1 \\
    \end{pmatrix}=\frac{1}{\sqrt{2}}\begin{pmatrix}
      \quad 1 \\
      -j \\
    \end{pmatrix}=\frac{1}{\sqrt{2}}\left(-\iota_A-j o_A\right).
\end{equation}

Then, the contraction gives, both numerically and formally, the voltage:
\begin{equation}\label{}
    V=\psi_A \eta^A=\psi_0\eta^0+\psi_1\eta^1=\frac{1}{2}\left(-\iota_A o^A+ o_A\iota^A \right)=1.
\end{equation}
For the spinor representation of the corresponding LHC transmitted  wave, under Graves' convention, whereby the representations differ for counterpropagating wave we use the primed frame.  First, lower the index of the antenna height, then apply the priming rule:
\begin{eqnarray}
  \eta^A &\rightarrow & \eta_A=\frac{1}{\sqrt{2}}\left(o_A + j\iota_A \right) \quad \rightarrow \\
   &\rightarrow & \psi_{\tilde{A}}=\frac{1}{\sqrt{2}}\left(o_{\tilde{A}} + j\iota_{\tilde{A}} \right)= \nonumber\\
   & \quad & \quad \:\:\: =\frac{1}{\sqrt{2}} \left[\begin{pmatrix}
      1 \\
      0 \\
    \end{pmatrix}+j\begin{pmatrix}
      0 \\
      1 \\
    \end{pmatrix}\right]=\frac{1}{\sqrt{2}}\begin{pmatrix}
      1 \\
      j \\
    \end{pmatrix}. \nonumber
\end{eqnarray}

As expected, the same handedness circular polarization wave has apparently the \emph{conjugate} representation, which arises through geometric and linear manipulations alone.  The same exercise can be repeated trivially for the case of RHC polarization, and one can consider the case of any linear polarization using real combinations of the spinor basis elements. The main difference in our approach is to regard the received wave as primary, and the transmitted wave as the \emph{reversed} direction. Conventionally, the transmitted wave has been treated in a sense as the primary one, the main justification being that it could be \emph{identified} with the antenna state.  Rigorous application of the formalism shows, on account of the tensor properties of the Green's dyadic that we can only obtain the componentwise  equality of the polarization state vectors by realizing Graves' directional wave vector formalism in terms of placing the antenna height vector and transmit field in different spin frames.  The received wave $\psi_A$ (described as a covariant spinor) lies in the dual space (i.e. in the same spin frame) as the antenna height $\eta^A$ (described as a contravariant), which it must to keep the scalar received voltage invariant.

\subsection{Basis transformations}
Again, considering the case of the circular polarizations, we can use the results above to deduce the basis transformation matrices ab initio.  For (H,V)$\, \rightarrow\,$(L,R), we have, for the received wave:
\begin{eqnarray}
  && \mbox{LHC: } \frac{1}{\sqrt{2}}\begin{pmatrix}
      \quad 1 \\
      -j \\
    \end{pmatrix} \quad \rightarrow \quad \begin{pmatrix}
      1 \\
      0 \\
    \end{pmatrix}  \\
   && \mbox{RHC: } \frac{1}{\sqrt{2}}\begin{pmatrix}
      -j \\
      \quad 1 \\
    \end{pmatrix} \quad \rightarrow \quad \begin{pmatrix}
      0 \\
      1 \\
    \end{pmatrix} \nonumber
\end{eqnarray}

\begin{equation}
   U_{\text{\tiny{R}}}=\frac{1}{\sqrt{2}}\begin{pmatrix}
      1 & j \\
      j &  1 \\
    \end{pmatrix}.
\end{equation}

While, for the transmitted wave,
\begin{eqnarray}
  && \mbox{LHC: } \frac{1}{\sqrt{2}}\begin{pmatrix}
       1 \\
      j \\
    \end{pmatrix} \quad \rightarrow \quad \begin{pmatrix}
      1 \\
      0 \\
    \end{pmatrix}  \\
   && \mbox{RHC: } \frac{1}{\sqrt{2}}\begin{pmatrix}
      j \\
       1 \\
    \end{pmatrix} \quad \rightarrow \quad \begin{pmatrix}
      0 \\
      1 \\
    \end{pmatrix} \nonumber
\end{eqnarray}

\begin{equation}
   U_{\text{\tiny{T}}}=\frac{1}{\sqrt{2}}\begin{pmatrix}
      \quad 1 & -j \\
      -j & \quad 1 \\
    \end{pmatrix}.
\end{equation}

We add suffices R, T here for clarity to denote where the transformation applies to the received or transmitted wave. The scattering matrix transformation, as Graves should have written it, then takes the form:
\begin{equation}\label{}
    \mathbf{E}_{\text{\tiny{R}}}=S\mathbf{E}_{\text{\tiny{T}}}\quad \rightarrow \quad \mathbf{E}'_{\text{\tiny{R}}}=U_{\text{\tiny{R}}}\mathbf{E}_{\text{\tiny{R}}}=U_{\text{\tiny{R}}} S (U_{\text{\tiny{T}}})^{-1}\mathbf{E}'_{\text{\tiny{T}}}
\end{equation}
which gives:
\begin{equation}\label{}
    S'=U_{\text{\tiny{R}}}\, S\, (U_{\text{\tiny{T}}})^{-1}=\frac{1}{2} \begin{pmatrix}
       1 & j \\
      j &  1 \\
    \end{pmatrix} S \begin{pmatrix}
      1 & j \\
      j &  1 \\
    \end{pmatrix}
\end{equation}
This is always automatically of the congruential form without any step involving a conjugation.

Given the basis transformation matrices for the received wave:
\begin{equation}\label{}
    U_\text{\tiny{R}}^{\text{\tiny{HV}}\rightarrow \text{\tiny{LR}}}=\frac{1}{\sqrt{2}}\begin{pmatrix}
       1 & j \\
      j &  1 \\
    \end{pmatrix} \quad U_\text{\tiny{R}}^{\text{\tiny{HV}}\rightarrow \pm 45^{\circ}}=\frac{1}{\sqrt{2}}\begin{pmatrix}
      1 & -1 \\
      1 & \quad 1 \\
    \end{pmatrix}
\end{equation}
a set of canonical examples of scattering matrices (sphere $S_{\text{\tiny{sph}}}$, diplane $S_{\text{\tiny{dipl}}}$ and left handed helix $S_{\text{\tiny{lhh}}}$) transforming under the congruential rule are reported in the following:
\begin{equation}\label{}
    \!\!S_{\text{\tiny{sph}}}^{\text{\tiny{HV}}}=\begin{pmatrix}
      1 & 0 \\
      0 & 1 \\
    \end{pmatrix}\quad \left\{\begin{array}{l}
                        \overset{U_\text{\tiny{R}}^{\text{\tiny{HV}}\rightarrow \text{\tiny{LR}}}}{\longrightarrow}  S_{\text{\tiny{sph}}}^{\text{\tiny{LR}}}=\begin{pmatrix}
      0 & j \\
      j & 0 \\
    \end{pmatrix} \\
                        \overset{U_\text{\tiny{R}}^{\text{\tiny{HV}}\rightarrow \pm 45^{\circ}}}{\longrightarrow} \!\!\! S_{\text{\tiny{sph}}}^{\pm 45^{\circ}}=\begin{pmatrix}
      1 & 0 \\
      0 & 1 \\
    \end{pmatrix}
                      \end{array} \right.
\end{equation}
\begin{equation}\label{}
    \!S_{\text{\tiny{dipl}}}^{\text{\tiny{HV}}}=\begin{pmatrix}
      1 & \quad 0 \\
      0 & -1 \\
    \end{pmatrix}\quad \left\{\begin{array}{l}
                        \overset{U_\text{\tiny{R}}^{\text{\tiny{HV}}\rightarrow \text{\tiny{LR}}}}{\longrightarrow}  S_{\text{\tiny{dipl}}}^{{\text{\tiny{LR}}}}=\begin{pmatrix}
      1 & \quad 0 \\
      0 & -1 \\
    \end{pmatrix} \\
                        \overset{U_\text{\tiny{R}}^{\text{\tiny{HV}}\rightarrow \pm 45^{\circ}}}{\longrightarrow} \!\!\! S_{\text{\tiny{dipl}}}^{\pm 45^{\circ}}=\begin{pmatrix}
      0 & 1 \\
      1 & 0 \\
    \end{pmatrix}
                      \end{array} \right.
\end{equation}

\begin{equation}\label{}
    S_{\text{\tiny{lhh}}}^{\text{\tiny{HV}}}=\frac{1}{\sqrt{2}}\begin{pmatrix}
      1 & \quad j \\
      j & -1 \\
    \end{pmatrix}\quad \left\{\begin{array}{l}
                        \overset{U_\text{\tiny{R}}^{\text{\tiny{HV}}\rightarrow \text{\tiny{LR}}}}{\longrightarrow}   S_{\text{\tiny{lhh}}}^{\text{\tiny{LR}}}=\sqrt{2}\begin{pmatrix}
      0 & \quad 0 \\
      0 & -1 \\
    \end{pmatrix} \\
                        \overset{U_\text{\tiny{R}}^{\text{\tiny{HV}}\rightarrow \pm 45^{\circ}}}{\longrightarrow} \!\!\! S_{\text{\tiny{lhh}}}^{\pm 45^{\circ}}=\frac{1}{\sqrt{2}}\begin{pmatrix}
      -j & 1 \\
      1 & j \\
    \end{pmatrix}
                      \end{array} \right.
\end{equation}

All the cases above involve unitary transformations, and there is no difference between the congruential rule and the result obtained by consimilarity. We now consider further examples, including two transformations that are unitary and one involving lossy propagation, in which the congruential rule still provides the correct answer, but where consimilarity throws up problems. We consider three simple targets: a sphere, a raindrop, and a $30^{\circ}$ canted dipole. The raindrop is considered to be oriented with symmetry axis vertical, and to have a copolar return in HH that is 3dB higher than in VV. The raw  scattering matrices in HV and LR basis are given in the following table.
\vspace{0.5cm}

\begin{table}[!t]
\renewcommand{\arraystretch}{1.9}
\caption{}
\label{table_example}
\centering
\begin{tabular}{|l|l|}
\hline
$S_{\text{\tiny{sph}}}^{\text{\tiny{HV}}}=\bigl(\begin{smallmatrix}
      1 & 0 \\
      0 & 1 \\
    \end{smallmatrix}\bigr) $ & $S_{\text{\tiny{sph}}}^{\text{\tiny{LR}}}=\bigl(\begin{smallmatrix}
      0 & j \\
      j & 0 \\
    \end{smallmatrix}\bigr) $ \\
\hline\hline
$S_{\text{\tiny{rd}}}^{\text{\tiny{HV}}}=\bigl(\begin{smallmatrix}
      S_{\text{\tiny{HH}}} & 0 \\
      0 & S_{\text{\tiny{VV}}} \\ \end{smallmatrix}\bigr)$ & $S_{\text{\tiny{rd}}}^{\text{\tiny{LR}}}=\bigl(\begin{smallmatrix}
      S_{\text{\tiny{HH}}} & 0 \\
      0 & S_{\text{\tiny{VV}}} \\ \end{smallmatrix}\bigr)$\\
\hline
\end{tabular}
\end{table}
\vspace{0.5cm}

\subsubsection{Doppler shift}

Consider the target to be moving with a recessional radian Doppler frequency, $\omega_d$. Then, as a time dependent  scattering matrix, we can express the scattering matrix in each case as equivalent to a time-dependent basis change (the antennas receding from the target): $S'=USU^T$ for example in the case of the raindrop,
\begin{eqnarray}\label{}
    S'&=&\begin{pmatrix}
      \mbox{e}^{\,j\,\omega_d t} & 0 \\
      0 & \mbox{e}^{\,j\,\omega_d t} \\
    \end{pmatrix}\begin{pmatrix}
      S_{\text{\tiny{HH}}} & 0 \\
      0 & S_{\text{\tiny{VV}}} \\
    \end{pmatrix}\begin{pmatrix}
      \mbox{e}^{\,j\,\omega_d t} & 0 \\
      0 & \mbox{e}^{\,j\,\omega_d t} \\
    \end{pmatrix} \nonumber \\
    &=& \begin{pmatrix}
      S_{\text{\tiny{HH}}}\,\mbox{e}^{2j\,\omega_d t} & 0 \\
      0 & S_{\text{\tiny{VV}}} \,\mbox{e}^{2j\,\omega_d t} \\
    \end{pmatrix}.
\end{eqnarray}
The well known double Doppler shift appears naturally in this prescription.  In the consimilarity representation, this comes about by the circuitous route of inverting and conjugating the return Doppler factor.

\subsubsection{Raindrop target in presence of lossy propagation factor}

European weather radars typically operate at C-band where absorption losses cannot be ignored.  Typically multiple scattering effects can be neglected, and the dominant effect is of coherent forward scattering with a mean attenuation (which we will here absorb into an overall scaling) and a differential attenuation $\Delta\tau$, and differential propagation phase $\Delta\varphi$ \cite{bringi}. Examples of this kind are not restricted to weather radars. Targets embedded in vegetation will also be modified by the anisotropy of scattering which may according to frequency involve both refractive and lossy characteristics of an effective medium. Now for the conguential rule we have, applying the transmission matrix $T$ to the received wave, and its transpose to the transmitted wave,
\begin{eqnarray}\label{}
    S'&=& TST^T= \nonumber\\
    &=&\begin{pmatrix}
      \mbox{e}^{\alpha} & 0 \\
      0 & \mbox{e}^{-\alpha} \\
    \end{pmatrix} \begin{pmatrix}
      S_{\text{\tiny{HH}}} & 0 \\
      0 & S_{\text{\tiny{VV}}} \\
    \end{pmatrix}\begin{pmatrix}
      \mbox{e}^{\alpha} & 0 \\
      0 & \mbox{e}^{-\alpha} \\
    \end{pmatrix} =\nonumber \\
    &=& \begin{pmatrix}
      S_{\text{\tiny{HH}}}\,\text{e}^{2\alpha} & 0 \\
      0 & S_{\text{\tiny{VV}}} \,\text{e}^{-2\alpha} \\
    \end{pmatrix}= \nonumber \\
    &=&\begin{pmatrix}
      S_{\text{\tiny{HH}}}\,\text{e}^{-\Delta\tau+j\Delta\varphi} & 0 \\
      0 & S_{\text{\tiny{VV}}} \,\text{e}^{\Delta\tau-j\Delta\varphi} \\
    \end{pmatrix},
\end{eqnarray}
where $\alpha=\frac{-\Delta\tau+j\Delta\varphi}{2}$.
But if we applied consimilarity the result would be:
\begin{eqnarray}\label{}
    S'&=& T S\bar{T}^{-1}=\nonumber \\
    &=&\begin{pmatrix}
      \mbox{e}^{\alpha} & 0 \\
      0 & \mbox{e}^{-\alpha} \\
    \end{pmatrix} \begin{pmatrix}
      S_{\text{\tiny{HH}}} & 0 \\
      0 & S_{\text{\tiny{VV}}} \\
    \end{pmatrix}\begin{pmatrix}
      \mbox{e}^{-\bar{\alpha}} & 0 \\
      0 & \mbox{e}^{\bar{\alpha}} \\
    \end{pmatrix} =\nonumber \\
    &=& \begin{pmatrix}
      S_{\text{\tiny{HH}}}\,\text{e}^{\alpha-\bar{\alpha}} & 0 \\
      0 & S_{\text{\tiny{VV}}} \,\text{e}^{\bar{\alpha}-\alpha} \\
    \end{pmatrix}= \nonumber \\
    &=&\begin{pmatrix}
      S_{\text{\tiny{HH}}}\,\text{e}^{j\Delta\varphi} & 0 \\
      0 & S_{\text{\tiny{VV}}} \,\text{e}^{-j\Delta\varphi} \\
    \end{pmatrix}.
\end{eqnarray}

This predicts that the relative amplitudes are unchanged, that is, that the return path compensates for the losses.  Experimentally, of course, this is not what is observed.  While $S_{\text{\tiny{HH}}}$ in this context is usually larger than $S_{\text{\tiny{VV}}}$, the differential attenuation is positive (i.e. more for HH than VV) and the ratio  $|\frac{S_{\text{\tiny{HH}}}}{S_{\text{\tiny{VV}}}}|$ as well as the absolute values is normally diminished when attenuation is present. If we had included the mean attenuation within the propagation matrix, this also would have been compensated on the return path.  For a fairly large raindrop of equivolume diameter 5 mm at 5.6 GHz, the scattering amplitudes are (in units of cm) $S_{\text{\tiny{HH}}}=(0.02021-j\,0.01044)$, $S_{\text{\tiny{VV}}}=(0.011885-j\,0.005359)$. In the case of not untypical propagation factors with a one-way differential attenuation of 0.24 dB, and a one-way integrated differential phase of $15^{\circ}$, the modified scattering measured matrices of the raindrop $S'_{\text{\tiny{rd}}}$ and unit dipole $S'_{\text{\tiny{dp}}}$ are (excluding the effects of the mean propagation effects)
\begin{equation}\label{}
    S'_{\text{\tiny{rd}}}=\begin{pmatrix}
      0.02162-j\,0.00462 & 0 \\
      0 & 0.01340-j\,0.0002 \\
    \end{pmatrix}
\end{equation}
\begin{equation}\label{}
    S'_{\text{\tiny{dp}}}=\begin{pmatrix}
      0.7047+j\,0.1888 & 0.433 \\
      0.433 & 0.2482+j\,0.0665 \\
    \end{pmatrix}.
\end{equation}

For the raindrop, na\"{i}ve application of consimilarity would have resulted in an error of nearly half a decibel in the differential reflectivity factor  $Z_{DR}=|\frac{S_{\text{\tiny{HH}}}}{S_{\text{\tiny{VV}}}}|^2$.

%% file: GP0Conclusion.tex

Graves' conception of directional wave vectors has been realized here using spinor representations, without recourse to physically incorrect complex conjugations. This respects the linearity of the scattering process, and there is no requirement for time-reversal symmetry, as shown by examples given where radar signals suffer (polarization dependent) attenuation.
The spinor priming operation provides the alternate representation for the counter-propagating wave.  Again, the analysis clears up an anomaly in that it has generally been presumed that the  transmitted electric field and antenna height can be identified up to dimensional or scale factors, with the return wave belonging to the \emph{opposite} directional wave space.  A critical understanding of the role of the Green's dyadic which is Euclidean invariant but not unitarily invariant shows that this axiomatic start point is in fact invalid; reliance on Cartesian coordinates in a complexified Euclidean space disguises the true transformational properties of the objects involved.  As Mieras correctly pointed out long ago \cite{mieras}, the basis invariance of the voltage equation is fundamental.  From there, it follows that the received field and antenna height transform reciprocally.  These belong to the same spin-frame representation (one covariant, the other contravariant) while the transmitted field in terms of Graves' directional wave formalism belongs to the alternative frame.  The idea of distinguishing covariant and contravariant objects is never required in Euclidean vector spaces, because the metric is the diagonal unit matrix preserved under (real) rotations and reflections - the isometries. Once unitary transformations are introduced this is no longer so.  Nor can one simply evade the problem by resorting to a Hermitian metric, because these map vectors from a pair of conjugate spaces to scalars, and this is inappropriate to the problem in hand.  Jones vectors were adopted in radar from optics, where there is practically no need to have a representation for antennas.  Without any specific provision in the formalism for their representation, the notion of their need to transform reciprocally with respect to fields has never fully crystallized because of the confusing counterfactual, that their polarization vector is to be identified with the transmitted field.  As our analysis showed, their componentwise equality arises from an appropriate choice of the alternate spin frame, but they are not formally identical from the geometric point of view.  Use of spinor notation forces us to realize that antennas and fields must exist in dual spaces, and provides the formal machinery for handling them consistently.
Graves' formulation refers to the scattering matrix in the way that it has generally been understood as the operator mediating between transmit and received field components. As we showed, and as Graves intended this operator has a domain in the space of outward propagating polarization states, and a range in the state of inward propagating states.  It clearly makes no sense for such scattering matrices to be concatenated, so they do not formally belong to a ring of matrix algebra, and one would never have expected that they would transform under similarity.  The congruential rule is not obvious either, however.  By contrast, the representation of the voltage equation as a voltage form (\ref{Vo}) naturally does, because it maps two antenna states in the same spin frame to a scalar. For this reason we would advocate that the doubly covariant voltage form be the preferred representation in radar, particularly for backscatter, since the symmetry that follows from reciprocity is immediately apparent, and equally, the congruential rule is naturally explained. It has been argued by some that the field operator form is in a sense more fundamental, but at least for BSA Graves' directional wave formalism requires quite involved artificial constructions to be expressed with mathematical and physical consistency.  We would counter that voltage form is what is actually measured; fields can only be inferred from measurements. There are several more reasons to favour the representations of both Sinclair and Kennaugh scattering matrices as bilinear forms that we hope to present in further publications.
Spinor formalism has played a crucial role in clarifying fundamental principles in polarimetric problems from both a mathematical and physical standpoint.  Existing formalism relies on the much more generally familiar Euclidean vector concepts but is inadequate for making the important distinctions that are required for the rigorous framework that polarimetry requires if we are to get the most out of exploring the subtleties of vector scattering that current and future technology will offer.  The adoption of spinor formalism would be as close to a paradigm shift as radar polarimetry has seen since it was pioneered, but it seems to be the ideal vehicle for presenting Geometric Polarimetry as a fresh and powerful means of problem description and solving.

%% file: GP0Appendix.tex

Let us consider first the spinor basis as in (\ref{contraspinbasis}) and (\ref{covariantspinbasis}) in its contravariant $\{o^A, \iota^A\}$ and covariant $\{o_A,\iota_A\}$ forms:
\begin{equation}\label{}
    o^A=  \begin{pmatrix} 
      o^0 \\
      o^1 \\
    \end{pmatrix}= \begin{pmatrix}
      1 \\
      0 \\
    \end{pmatrix}, \quad \iota^A= \begin{pmatrix}
      \iota^0 \\
      \iota^1 \\
    \end{pmatrix}=\begin{pmatrix}
      0 \\
      1 \\
    \end{pmatrix},
\end{equation}
\begin{equation}\label{covariantspinbasisI}
    o_B=o^A \varepsilon_{AB}=\begin{pmatrix}
      0 \\
      1 \\
    \end{pmatrix}, \quad \iota_B=\iota^A \varepsilon_{AB}=\begin{pmatrix}
      -1 \\
      \quad \!0 \\
    \end{pmatrix}.
\end{equation}
The spinor $M_A^{\,\,\,\,B}$ in (\ref{eta}) describing linear operations on covariant spinors can be constructed from a basis for all these matrices. This basis is easily built from the outer products of the basis spinors $o^A$, $\iota^A$ as:
\begin{eqnarray}
  o_A o^B &=& \begin{pmatrix}
              o_0\, o^0 & o_0\, o^1 \\
              o_1\, o^0 & o_1\, o^1 \\
            \end{pmatrix}=\begin{pmatrix}
                            0 & 0 \\
                            1 & 0 \\
                          \end{pmatrix} \\
  o_A \iota^B &=& \begin{pmatrix}
                            0 & 0 \\
                            0 & 1 \\
                          \end{pmatrix}  \\
  \iota_A o^B &=& \begin{pmatrix}
                            -1 & 0 \\
                            \quad\!0 & 0\\
                          \end{pmatrix}  \\
  \iota_A \iota^B &=& \begin{pmatrix}
                            0 & -1 \\
                            0 & \quad\!0 \\
                          \end{pmatrix}.
\end{eqnarray}
The spinor $M_A^{\,\,\,\,B}$ can be expressed as:
\begin{equation}\label{}
    M_A^{\,\,\,\,B}=\gamma\, o_A o^B+\delta\, o_A \iota^B-\alpha\, \iota_A o^B-\beta\, \iota_A \iota^B,
\end{equation}
which using the matrix form of the outer products becomes
\begin{equation}\label{}
    M_A^{\,\,\,\,B}=\begin{pmatrix}
                            \alpha & \beta \\
                            \gamma & \delta \\
                          \end{pmatrix}.
\end{equation}
The basis spinor $o_A$ and $\iota_A$ will transform in the new spinors $\kappa_A$ and $\lambda_A$ as in the following:
\begin{equation}\label{}
    \kappa_A=M_A^{\,\,\,\,B}o_B=\begin{pmatrix}
                                     \beta \\
                                     \delta \\
                                   \end{pmatrix}, \quad \kappa^A=\begin{pmatrix}
                                                                      \quad\!\delta \\
                                                                      -\beta \\
                                                                    \end{pmatrix}
\end{equation}
\begin{equation}\label{}
    \lambda_A=M_A^{\,\,\,\,B}i_B=\begin{pmatrix}
                                     -\alpha \\
                                     -\gamma \\
                                   \end{pmatrix}, \quad \lambda^A=\begin{pmatrix}
                                                                      -\gamma \\
                                                                      \quad\!\alpha \\
                                                                    \end{pmatrix},
\end{equation}
and their inner product is:
\begin{equation}\label{}
    \kappa_A\lambda^A=\delta \alpha-\beta \gamma.
\end{equation}
Using instead the new spinor basis $o^{\tilde{A}}$ and $\iota^{\tilde{A}}$ obtained with the priming operation in (\ref{otilde})
\begin{equation}\label{}
    o^{\tilde{A}}=-\iota^A=\begin{pmatrix}
                      \quad \!0 \\
                      -1 \\
                    \end{pmatrix}, \quad \iota^{\tilde{A}}=o^A=\begin{pmatrix}
                      1 \\
                      0 \\
                    \end{pmatrix}
\end{equation}
and their covariant form
\begin{equation}\label{}
    o_{\tilde{B}}=o^{\tilde{A}}\varepsilon_{\tilde{A}\tilde{B}}=\begin{pmatrix}
                      1 \\
                      0 \\
                    \end{pmatrix}, \quad \iota_{\tilde{B}}=\iota^{\tilde{A}}\varepsilon_{\tilde{A}\tilde{B}}=\begin{pmatrix}
                      0  \\
                      1 \\
                    \end{pmatrix}.
\end{equation}
Their outer products are:
\begin{eqnarray}
  o_{\tilde{A}} o^{\tilde{B}}  &=& \begin{pmatrix}
                            0 & -1 \\
                            0 & \quad\!0 \\
                          \end{pmatrix} \\
  o_{\tilde{A}} \iota^{\tilde{B}} &=& \begin{pmatrix}
                            1 & 0 \\
                            0 & 0 \\
                          \end{pmatrix}  \\
  \iota_{\tilde{A}} o^{\tilde{B}} &=& \begin{pmatrix}
                            0 & \quad\!0 \\
                            0 & -1\\
                          \end{pmatrix}  \\
  \iota_{\tilde{A}} \iota^{\tilde{B}} &=& \begin{pmatrix}
                            0 & 0 \\
                            1 & 0 \\
                          \end{pmatrix}.
\end{eqnarray}
The spinor $M_A^{\,\,\,\,B}$ can be expressed as:
\begin{equation}\label{}
    M_{\tilde{A}}^{\,\,\,\,{\tilde{B}}}=\gamma\, o_{\tilde{A}} o^{\tilde{B}}+\delta\, o_{\tilde{A}} \iota^{\tilde{B}}-\alpha\, \iota_{\tilde{A}} o^{\tilde{B}}-\beta\, \iota_{\tilde{A}} \iota^{\tilde{B}},
\end{equation}
which using the matrix form of the outer products becomes
\begin{equation}\label{}
    M_{\tilde{A}}^{\,\,\,\,{\tilde{B}}}=\begin{pmatrix}
                            \delta & -\gamma \\
                            -\beta & \alpha \\
                          \end{pmatrix}.
\end{equation}
The basis spinor $o_{\tilde{A}}$ and $\iota_{\tilde{A}}$ will transform in the new spinors $\kappa_{\tilde{A}}$ and $\lambda_{\tilde{A}}$:
\begin{equation}\label{}
    \kappa_{\tilde{A}}=M_{\tilde{A}}^{\,\,\,\,{\tilde{B}}}o_{\tilde{B}}=\begin{pmatrix}
                                     \quad\!\delta \\
                                     -\beta \\
                                   \end{pmatrix}, \quad \kappa^{\tilde{A}}=\begin{pmatrix}
                                                                      -\beta \\
                                                                      -\delta \\
                                                                    \end{pmatrix}
\end{equation}
\begin{equation}\label{}
    \lambda_{\tilde{A}}=M_{\tilde{A}}^{\,\,\,\,{\tilde{B}}}\iota_{\tilde{B}}=\begin{pmatrix}
                                     -\gamma \\
                                     \quad\!\alpha \\
                                   \end{pmatrix}, \quad \lambda^{\tilde{A}}=\begin{pmatrix}
                                                                      \alpha \\
                                                                      \gamma \\
                                                                    \end{pmatrix},
\end{equation}
and their inner product is:
\begin{equation}\label{}
    \kappa_{\tilde{A}}\lambda^{\tilde{A}}=\delta \alpha-\beta \gamma.
\end{equation}
The null basis vectors $\mathbf{l}$, $\mathbf{n}$, $\mathbf{m}$ and $\bar{\mathbf{m}}$ in (\ref{lnm:spinor}) will transform under the priming operation in:
\begin{equation}\label{}
    \begin{array}{l}
      \tilde{\mathbf{l}} \;\,\,\, \rightarrow \; o^{\tilde{A}} o^{\tilde{A}'}\equiv \begin{pmatrix}
        0 & 0 \\
        0 & 1 \\
      \end{pmatrix}\\
       \tilde{\mathbf{n}} \; \,\rightarrow \; \iota^{\tilde{A}} \iota^{\tilde{A}'}\,\equiv \begin{pmatrix}
        1 & 0 \\
        0 & 0 \\
      \end{pmatrix} \\
       \tilde{\mathbf{m}} \; \rightarrow \; o^{\tilde{A}} \iota^{\tilde{A}'}\equiv \begin{pmatrix}
        \quad\!0 & 0 \\
        -1 & 0 \\
      \end{pmatrix} \\
       \tilde{\bar{\mathbf{m}}} \; \rightarrow \; \iota^{\tilde{A}} o^{\tilde{A}'}\equiv \begin{pmatrix}
        0 & -1 \\
        0 & \quad\!0 \\
      \end{pmatrix} \\
    \end{array},
\end{equation}
namely $\tilde{\mathbf{l}}\; \rightarrow \; \mathbf{n}$, $\tilde{\mathbf{n}}\; \rightarrow \; \mathbf{l}$, $\tilde{\mathbf{m}}\; \rightarrow \; -\bar{\mathbf{m}}$ and $\tilde{\bar{\mathbf{m}}}\; \rightarrow \; -\mathbf{m}$ as in (\ref{lnmpriming}).

%% file: GEOpolarimetry0.bbl
\begin{thebibliography}{10}
\providecommand{\url}[1]{#1}
\csname url@samestyle\endcsname
\providecommand{\newblock}{\relax}
\providecommand{\bibinfo}[2]{#2}
\providecommand{\BIBentrySTDinterwordspacing}{\spaceskip=0pt\relax}
\providecommand{\BIBentryALTinterwordstretchfactor}{4}
\providecommand{\BIBentryALTinterwordspacing}{\spaceskip=\fontdimen2\font plus
\BIBentryALTinterwordstretchfactor\fontdimen3\font minus
  \fontdimen4\font\relax}
\providecommand{\BIBforeignlanguage}[2]{{%
\expandafter\ifx\csname l@#1\endcsname\relax
\typeout{** WARNING: IEEEtran.bst: No hyphenation pattern has been}%
\typeout{** loaded for the language `#1'. Using the pattern for}%
\typeout{** the default language instead.}%
\else
\language=\csname l@#1\endcsname
\fi
#2}}
\providecommand{\BIBdecl}{\relax}
\BIBdecl

\bibitem{galletti2008}
M.~Galletti, D.~Bebbington, M.~Chandra, and T.~Borner, ``Measurement and
  characterization of entropy and degree of polarization of weather radar
  targets,'' \emph{IEEE Trans. on Geoscience and Remote Sensing}, vol. 46, no.
  10, pp. 3196--3207, 2008.

\bibitem{galletti2011}
M.~Galletti, D.~Zrnic, V.~Melnikov, and R.~Doviak, ``Degree of polarization at
  horizontal transmit: Theory and applications for weather radar,'' \emph{IEEE
  Trans. on Geoscience and Remote Sensing}, vol. PP, no. 99, pp. 1--11, 2011.

\bibitem{paladini}
R.~Paladini, M.~Martorella, and F.~Berizzi, ``Classification of man-made
  targets via invariant coherency-matrix eigenvector decomposition of
  polarimetric {S}{A}{R}/{I}{S}{A}{R} images,'' \emph{IEEE Trans. on Geoscience
  and Remote Sensing}, vol. 49, no.8, pp. 3022--3034, 2011.

\bibitem{luneburgreply}
E.~L{\"u}neburg, ``Comments on the '{S}pecular null polarization theory', by
  {J}.{C}. {H}ubbert,'' \emph{IEEE Trans. on Geoscience and Remote Sensing},
  vol. 35, no. 4, pp. 1070--1071, 1997.

\bibitem{hubbertreply}
J.~Hubbert, ``Reply to '{C}omments on {T}he specular null polarization theory',
  by {E}. {L}{\"u}neburg,'' \emph{IEEE Trans. on Geoscience and Remote
  Sensing}, vol. 35, no.4, pp. 1071--1072, 1997.

\bibitem{kostinski}
A.~Kostinski and W.-M. Boerner, ``On the foundations of radar polarimetry,''
  \emph{IEEE Trans. on Antenna and Propagation}, vol.~34, pp. 1395--1404, 1986.

\bibitem{mieras}
H.~Mieras, ``Comments on '{O}n the foundations of radar polarimetry', by
  {A}.{B}. {K}ostinski and {W}-{M}. {B}oerner,'' \emph{IEEE Trans. on Antenna
  and Propagation}, vol.~34, pp. 1470--1471, 1986.

\bibitem{kostinskiR}
A.~Kostinski and W.-M. Boerner, ``Reply to '{C}omments on {O}n the foundations
  of radar polarimetry' by {H}. {M}ieras,'' \emph{IEEE Trans. on Antenna and
  Propagation}, vol.~34, pp. 1471--1473, 1986.

\bibitem{hubbertbringi}
J.~Hubbert and V.~Bringi, ``Specular null polarization theory: Applications to
  radar meteorology,'' \emph{IEEE Trans. on Geoscience and Remote Sensing},
  vol. 34, no.4, pp. 859--873, 1996.

\bibitem{graves}
C.~Graves, ``Radar polarization power scattering matrix,'' \emph{Proceedings of
  I.R.E.}, vol.~44, pp. 248--252, February 1956.

\bibitem{cloude:liegroups}
S.~Cloude, ``Ch. 2 - {L}ie groups in electromagnetic wave propagation and
  scattering,'' in \emph{Electromagnetic Symmetry}, C.~Baum and H.~Kritikos,
  Eds.\hskip 1em plus 0.5em minus 0.4em\relax Washington: Taylor and Francis,
  1995, pp. 91--142.

\bibitem{cloude:vector}
------, ``Vector {E}{M} scattering theory for polarimetric {S}{A}{R} image
  interpretation,'' in \emph{Direct and Inverse Electromagnetic Scattering},
  ser. Pitman Research Notes in Mathematics Series 361, H.~Serbest and
  S.~Cloude, Eds.\hskip 1em plus 0.5em minus 0.4em\relax Harlow, UK: Addison
  Wesley Longman, 1996, pp. 217--230.

\bibitem{torres}
G.~Torres{ d}el{ C}astillo and I.~Rubalcava{ G}arc\'{i}a, ``The {J}ones vector
  as a spinor and its representation on the {P}oincar\'{e} sphere,''
  \emph{Revista Mexicana de Fisica}, vol.~57, pp. 406--413, 2011.

\bibitem{carrea:igarss}
L.~Carrea and G.~Wanielik, ``Polarimetric {S}{A}{R} processing using the polar
  decomposition of the scattering matrix,'' in \emph{Proc. of IGARSS 2001 -
  International Geoscience and Remote Sensing}, Sydney, Australia, 9-13 July
  2001.

\bibitem{souyris}
J.-C. Souyris and C.~Tison, ``Polarimetric analysis of bistatic {S}{A}{R}
  images from polar decomposition: a quaternion approach,'' \emph{IEEE Trans.
  on Geoscience and Remote Sensing}, vol. 45, no.9, pp. 2701--2714, 2007.

\bibitem{huynen}
J.~Huynen, ``Phenomenological theory of radar targets,'' in
  \emph{Electromagnetic Scattering}, P.~Uslenghi, Ed.\hskip 1em plus 0.5em
  minus 0.4em\relax Academic Press New York, 1978, pp. 653--712.

\bibitem{bebbington:eusar}
D.~Bebbington, E.~Krogager, and M.~Hellmann, ``Vectorial generalization of
  target helicity,'' in \emph{3$^{rd}$ {E}uropean {C}onference on {S}ynthetic
  {A}perture {R}adar ({E}u{S}{A}{R})}, May 2000, pp. 531--534.

\bibitem{luneburg}
E.~L{\"u}neburg, ``Aspects of radar polarimetry,'' \emph{Elektrik-Turkish
  Journal of Electrical Engineering and Computer Science}, vol. 10, no.2, pp.
  219--243, 2002.

\bibitem{jonesI}
R.~Jones, ``A new calculus for the treatment of optical systems {I}:
  Description and discussion,'' \emph{J. Opt. Soc. Am.}, vol.~31, pp. 488--493,
  1941.

\bibitem{sinclair}
G.~Sinclair, ``The transmission and reception of elliptically polarized
  waves,'' \emph{Proc. of {I}{R}{E}}, vol.~38, pp. 148--151, 1950.

\bibitem{ieee83}
IEEE, \emph{Standard {D}efinitions of Terms for Antennas ({I}{E}{E}{E} {S}td
  145-1983)}, 1983.

\bibitem{mottI}
H.~Mott, \emph{Polarization in Antennas and Radar}.\hskip 1em plus 0.5em minus
  0.4em\relax John Wiley and Sons, 1986.

\bibitem{bebbington:invariance}
D.~Bebbington, L.~Carrea, and G.~Wanielik, ``Applications of geometric
  polarization to invariance properties in bistatic scattering,''
  \emph{Advances in Radio Science}, vol.~3, pp. 421--425, 2005.

\bibitem{luneburg:revision}
E.~L{\"u}neburg, ``Radar polarimetry, a revision of basic concepts,'' in
  \emph{Direct and Inverse Electromagnetic Scattering}, ser. Pitman Research
  Notes in Mathematics Series 361, H.~Serbest and S.~Cloude, Eds.\hskip 1em
  plus 0.5em minus 0.4em\relax Harlow, UK: Addison Wesley Longman, 1996, pp.
  257--275.

\bibitem{hornjohnson}
R.~Horn and C.~Johnson, \emph{Matrix Analysis}.\hskip 1em plus 0.5em minus
  0.4em\relax Cambridge University Press, 1985.

\bibitem{luneburg:finalreport}
E.~L{\"u}neburg, ``Final report phase {I}: Foundation of the mathematical
  theory of polarimetry,'' {O}{N}{R} {C}ontract {N}0001400{-}{M}0152, {E}{M}{L}
  {C}onsultants, 82234 Wessling, Germany, Tech. Rep.

\bibitem{mccormick:priciples}
G.~McCormick and A.~Hendry, ``Principles for the radar determination of the
  polarization properties of precipitation,'' \emph{Radio Science}, vol. 10,
  no.4, pp. 421--434, 1975.

\bibitem{mccormick:prop}
G.~McCormick, ``Propagation through a precipitation medium: Theory and
  measurement,'' \emph{IEEE Trans. on Antenna and Propagation}, vol. 23 no.2,
  pp. 266--269, 1975.

\bibitem{bebbington:correction}
D.~Bebbington, R.~McGuinness, and A.~Holt, ``Correction of propagation effects
  in {S}-band circular polarization-diversity radar,'' \emph{IEE Proceedings H
  134}, pp. 431--437, 1987.

\bibitem{akhmanov}
S.~Akhmanov and S.~Y. Nikitin, \emph{Physical Optics}.\hskip 1em plus 0.5em
  minus 0.4em\relax Oxford, UK: Clarendon Press, 1997, ch.~22.

\bibitem{nieto}
M.~Nieto-Vesperinas, \emph{Scattering and Diffraction in Physical
  Optics}.\hskip 1em plus 0.5em minus 0.4em\relax New York: J. Wiley \& Sons
  Inc., 1991, ch.~8.

\bibitem{alvarez}
J.-L. Alvarez{ P}erez, ``Coherence, polarization, and statistical independence
  in {C}loude-{P}ottier's radar polarimetry,'' \emph{IEEE Trans. Geoscience and
  Remote Sensing}, vol. 49, no.1, pp. 426--441, 2011.

\bibitem{post}
E.~Post, \emph{Formal Structure of Electromagnetics - General Covariance and
  Electromagnetics}.\hskip 1em plus 0.5em minus 0.4em\relax Dover Publications
  Inc., 1998.

\bibitem{cartan}
E.~Cartan, \emph{The Theory of Spinors}.\hskip 1em plus 0.5em minus 0.4em\relax
  Paris: Hermann, 1966.

\bibitem{woodhouse}
N.~Woodhouse, \emph{Geometric Quantization}.\hskip 1em plus 0.5em minus
  0.4em\relax Oxford Science Publications, Clarendon Press, 1991.

\bibitem{payne}
W.~Payne, ``Elementary spinor theory,'' \emph{American Journal of Physics},
  vol. 20 no. 5, pp. 253--261, 1952.

\bibitem{penrose}
R.~Penrose and W.~Rindler, \emph{Spinors and Space-Time}.\hskip 1em plus 0.5em
  minus 0.4em\relax Cambridge University Press, 1984, vol.~1.

\bibitem{pauli}
W.~Pauli, ``Zur {Q}uantenmechanik des magnetischen {E}lektrons,''
  \emph{Zeitschrift fuer Physik}, vol. 43 no. 9-10, pp. 601--623, 1927.

\bibitem{cloude:liegroups:journal}
S.~Cloude, ``{L}ie groups in electromagnetic wave propagation and scattering,''
  \emph{Journal of Electromagnetic Waves Applications}, vol. 6, no.8, pp.
  947--974, 1992.

\bibitem{bringi}
V.~Bringi and V.~Chandrasekhar, \emph{Polarimetric Doppler Weather
  Radar}.\hskip 1em plus 0.5em minus 0.4em\relax Cambridge University Press,
  2001, ch.~4.

\end{thebibliography}
